\documentclass[onecolumn,useAMS,usenatbib]{mn2e}
\usepackage{bm,graphicx,natbib}
\pdfoutput=1

\newcommand\aj{{AJ}}
\newcommand\apj{ApJ}
\newcommand\apjl{ApJ}
\newcommand\mnras{{MNRAS}}
\newcommand\aap{{A\&A}}

\newcommand\Rfold{R_{\rm fold}}
\newcommand\Rcusp{R_{\rm cusp}}

\newcommand\Acusp{A_{\rm cusp}}

\newcommand\bx   {\vect{\rm x}}
\newcommand\bby  {\vect{\rm y}}

\newcommand\bmo  {0}
\newcommand\bt   {\vect{\rm \theta}}
\newcommand\bu   {\vect{\rm u}}
\newcommand\grad{\vect{\nabla}}

\newcommand\ha   {\hat{a}}
\newcommand\hb   {\hat{b}}
\newcommand\hc   {\hat{c}}

\newcommand\vect[1]{{\mbox{\boldmath $#1$}}}
\newcommand\order[2]{\mathcal{O}\left(#1\right)^{#2}}
\newcommand\orderof[1]{\mathcal{O}\left(#1\right)}

\begin{document}

\title[Analytic Relations for Fold and Cusp Lenses]
{Analytic Relations for Magnifications and Time Delays in Gravitational
Lenses with Fold and Cusp Configurations}

\author[A. B. Congdon, C. R. Keeton and C. E. Nordgren]
{Arthur B. Congdon$^{1}$, Charles R. Keeton$^{1}$ and C. Erik Nordgren$^{2}$ \\
$^{1}$Department of Physics and Astronomy, Rutgers University,
136 Frelinghuysen Road, Piscataway, NJ 08854 USA; \\
acongdon, keeton@physics.rutgers.edu \\
$^{2}$nordgren@sas.upenn.edu}

\maketitle

\begin{abstract}

Gravitational lensing provides a unique and powerful probe of the mass distributions of distant galaxies.  Four-image lens systems with fold and cusp configurations have two or three bright images near a critical point. Within the framework of singularity theory, we derive analytic relations that are satisfied for a light source that lies a small but finite distance from the astroid caustic of a four-image lens. Using a perturbative expansion of the image positions, we show that the time delay between the close pair of images in a fold lens scales with the cube of the image separation, with a constant of proportionality that depends on a particular third derivative of the lens potential. We also apply our formalism to cusp lenses, where we develop perturbative expressions for the image positions, magnifications and time delays of the images in a cusp triplet. Some of these results were derived previously for a source asymptotically close to a cusp point, but using a simplified form of the lens equation whose validity may be in doubt for sources that lie at astrophysically relevant distances from the caustic. Along with the work of \citet{Keeton-fold}, this paper demonstrates that perturbation theory plays an important role in theoretical lensing studies.

\end{abstract}


\section{Introduction} \label{sec:intro}

Gravitational lensing, or the bending of light by gravity, offers an
exciting synergy between mathematics and astrophysics. Singularity
theory provides a powerful way to describe lensing near critical
points \citep{Schneider_lensing,Petters_book}, which turns out
to have important implications for astrophysics and the quest to
understand dark matter. Four-image lensed quasars can be broadly
classified, according to image geometry, as either folds, cusps or
crosses.  We are interested in folds and cusps, which occur when the
light source is close to the caustic curve, along which the lensing
magnification is infinite (see top row of Fig.~\ref{fig:u-theta-def}).
Fold lenses feature a close pair of bright images, while cusp lenses
have a close triplet of bright images (see bottom row of
Fig.~\ref{fig:u-theta-def}). By expanding the gravitational potential
of the lens galaxy in a Taylor series, one finds that the image
magnifications satisfy simple analytic relations, viz.
\begin{equation}
|\mu_A| - |\mu_B| \approx 0 \quad {\rm and} \quad |\mu_A| - |\mu_B| + |\mu_C|
\approx 0
\end{equation}
for fold pairs and cusp triplets, respectively \citep{Blandford_Narayan,
Mao_cusp, Schneider_cusp, Schneider_lensing, zakharov-magsum, Gaudi-fold, Gaudi-cusp,
Keeton-cusp, Keeton-fold}. For a fold pair, the two images have opposite parity, hence the negative sign. For a cusp triplet, the middle image (B) has opposite parity from the outer images (A and C). Note that in practice, one works with the image fluxes, which are directly observable, rather than the
magnifications, which are not. This leads to the equivalent relations:
\begin{equation} \label{eq:fluxratdef}
\Rfold \equiv \frac{F_A - F_B}{F_A + F_B} \approx 0 \quad {\rm and} \quad
\Rcusp \equiv \frac{F_A - F_B + F_C}{F_A + F_B + F_C} \approx 0 ,
\end{equation}
where the image flux $F_i$ is related to the source flux $F_0$ by $F_i = |\mu_i| F_0$.

Observationally, the fold and cusp relations are violated in several
lens systems \citep[e.g.,][]{Hogg_1422, Falco_0414, Keeton-shear-ellip,
Keeton-cusp, Keeton-fold}.  These so-called ``flux-ratio anomalies''
are taken as strong evidence that lens galaxies contain significant
small-scale structure \citep[e.g.,][]{Mao-flux, Metcalf-substruc,
Chiba_substruc}.  Since global modifications of the lens potential
(modeled with multipole terms, for example) cannot explain the
anomalous flux ratios \citep{Evans-multipole, Congdon-multipole, Yoo-1115, Yoo-rings},
it is generally believed that lens galaxies must contain substructure
in the form of mass clumps on a scale $\ga 10^6\,M_\odot$
\citep{Dobler-finite}.  Indeed, substructure is observed in at least
two lens systems, in the form of a dwarf galaxy near the main lens
galaxy \citep{Schechter-0414, McKean-2045}.  The substructure could
well be invisible, though; numerical simulations of structure
formation in the cold dark matter (CDM) paradigm predict that
galaxy dark matter halos are filled with hundreds of dark subhalos
\citep[e.g.,][]{Klypin-missing, Moore-substructure}.  The abundance
of flux-ratio anomalies in lensed radio sources implies that lens
galaxies contain $\sim$2\% (0.6--7\% at 90\% confidence) of their
mass in substructure \citep{Dalal-substruc}.  This may be somewhat
higher than the amount of CDM substructure
\citep[e.g.,][]{Mao-substructure}, although new higher-resolution
simulations are attempting to refine the theoretical predictions
and make more direct comparisons with lensing observations
\citep[e.g.,][]{Gao-substructure, Strigari-VL, Diemand-vialactea}.
While there is still work to be done along these lines, it is
clear that violations of the ideal fold and cusp relations provide
important constraints on the small-scale distribution of matter
in distant lens galaxies.

\citet{Keeton-cusp, Keeton-fold} pointed out that the ideal fold
and cusp relations only hold for a source asymptotically close
to the caustic.  If we want to use flux-ratio anomalies to
study small-scale structure in lens galaxies, it is vital that we
understand how much $\Rfold$ and $\Rcusp$ can deviate from zero
for realistic smooth lenses just because the source lies a small
but finite distance from the caustic.  To that end,
\citet{Keeton-cusp} used a Taylor series approach to demonstrate
that $\Rcusp = 0$ to lowest order, and used Monte Carlo simulations
to suggest that $\Rcusp \propto d^2$, where $d$ is the distance
between the two most widely separated cusp images.  In order to
derive the leading non-vanishing term in $\Rcusp$ analytically,
it would be necessary to extend the Taylor series to a higher order
of approximation. This has not been done before, since including
higher-order terms substantially complicates the analysis. Instead,
previous authors have made assumptions about which terms are
important and which terms can be neglected, in order to obtain an
analytically manageable problem.

As we shall see, perturbation theory provides a natural way to
overcome the difficulties of the Taylor series approach.
\citet{Keeton-fold} used perturbation theory to show that for fold
lenses, $\Rfold$ is proportional to the image separation $d_1$ of
the fold pair.  We extend their analysis in several important ways.
We derive the leading-order non-vanishing expression for $\Rcusp$
using perturbation theory. We also show, for the first time, how
the combination of singularity theory and perturbation theory can
be used to study {\em time delays} between lensed images for both
fold and cusp systems.  Working in analogy to flux ratios,
we derive time-delay relations for fold and cusp lenses, which we
will apply (in a forthcoming paper) to observed lenses in order to
identify ``time-delay anomalies.''  As time delays join flux ratios
in probing small-scale structure in lens galaxies
\citep[see][]{Keeton-tdel}, singularity theory and perturbation
theory once again provide the rigorous mathematical foundation.

\section{Mathematical Preliminaries \label{sec:coords}}

\begin{figure}
\begin{center}
\includegraphics[width=0.9\textwidth]{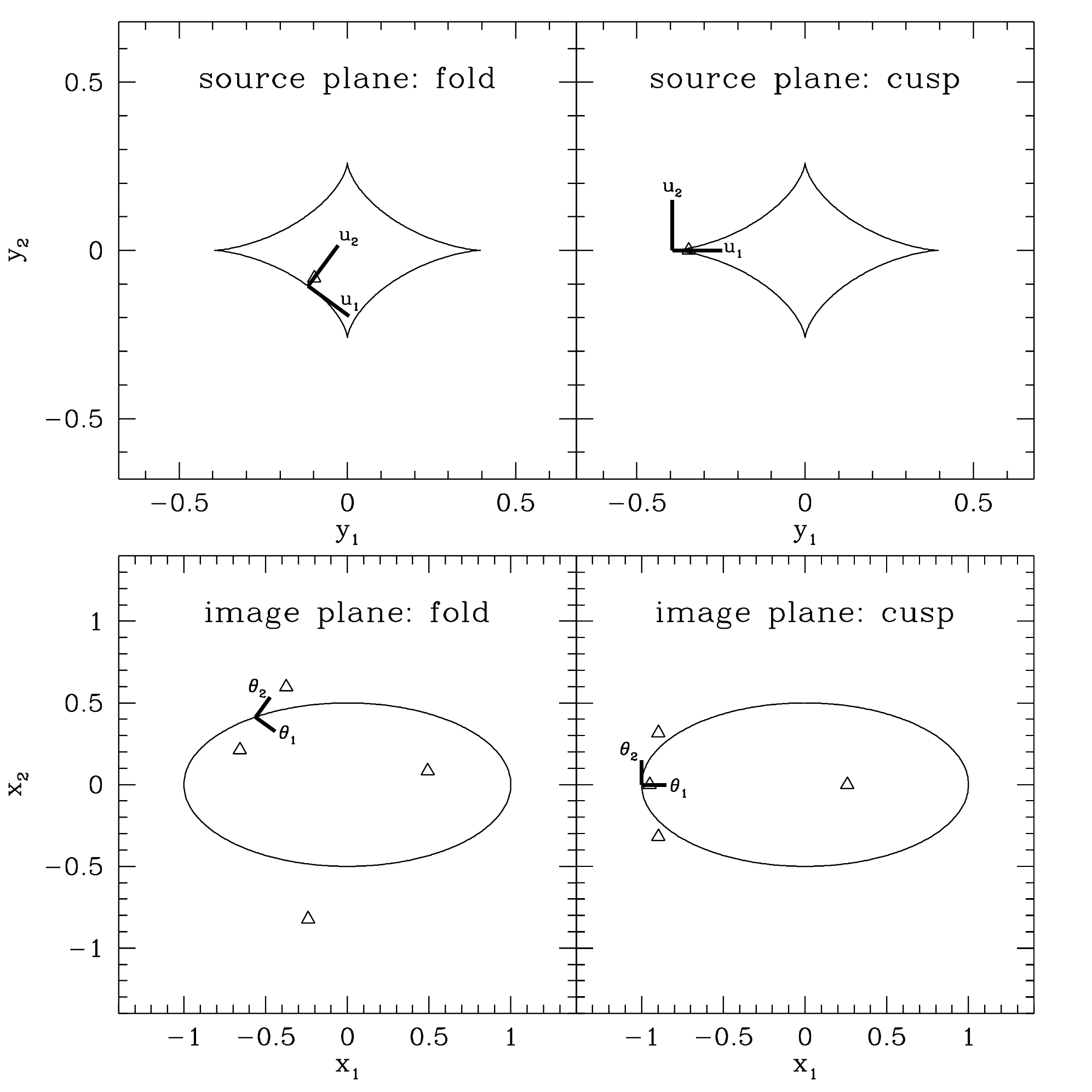}
\end{center}
\caption[Local orthogonal coordinates]{The curves show the caustic (top) and critical curve (bottom) for a
singular isothermal ellipsoid (SIE) lens with minor-to-major axis
ratio $q=0.5$.  The local orthogonal coordinates defined by the
rotation matrix ${\bf M}$ are indicated for a fold point (left) and a cusp
point (right). In the top two panels, a source (triangle) with position
$(y_1,y_2)$ measured from the center of the caustic (astroid) has
position $(u_1,u_2)$ in the rotated coordinates centered on a fold
point (left panel) or a cusp (right panel). In the bottom two panels, the four images (triangles) have
positions $(x_1,x_2)$ measured from the center of the critical curve
(ellipse) and positions $(\theta_1, \theta_2)$ in the rotated coordinates.
\label{fig:u-theta-def}}
\end{figure}

To study lensing of a source near a caustic, it is convenient to
work in coordinates centered at a point on the caustic. Nonspherical
lenses typically have two caustics. The ``radial'' caustic separates
regions in the source plane for which one (outside) and two (inside)
images are produced.\footnote{We are neglecting faint images
predicted to form near the centers of lens galaxies, because they
are highly demagnified and difficult to detect.}  Within the radial
caustic is the ``tangential'' caustic or astroid, which separates
regions in the source plane for which two (between the two caustics)
and four (inside the astroid) images are produced. We are interested
in four-image lenses, where the source is within the astroid. We
therefore make no further reference to the radial caustic. A typical
astroid is shown in Figure \ref{fig:u-theta-def} for a lens galaxy
modeled by a singular isothermal ellipsoid (SIE), which is commonly
used in the literature \citep[e.g.,][]{Kormann_SIE} and
appears to be quite a good model for real lens galaxies
\citep[e.g.,][]{Rusin-field-ellip, Koopmans-SLACS-III}.  It is
customary to define source-plane coordinates $(y_1,y_2)$ centered
on the caustic and aligned with its symmetry axes. However, for
fold and cusp configurations, where the source is a small distance
from the caustic, it is more natural to work in coordinates $(u_1,u_2)$
centered on the fold or cusp point. For convenience, we define the
$u_1$ axis tangent to (and the $u_2$ axis orthogonal to) the caustic
at that point. Transforming from the $(y_1,y_2)$ plane to the
$(u_1,u_2)$ plane requires a translation plus a rotation. To derive
this coordinate transformation, we follow the discussion in Appendix
A1 in \citet{Keeton-fold}, which summarizes the results of
\citet{Petters_book}.

We begin by considering the lens equation $\bby = \bx -
\grad\psi(\bx)$, which maps the image plane to the source plane. The
solutions to this equation give the image positions $\bx \equiv
(x_1,x_2)$ corresponding to a given source position $\bby \equiv
(y_1,y_2)$. The function $\psi(\bx)$ is the scaled
gravitational potential of the lens galaxy projected onto the lens
plane. A caustic is a curve along which the magnification is
infinite, i.e., $\det (\partial \bby / \partial \bx)= \mu^{-1} = 0$,
where $\partial \bby / \partial \bx$ is the Jacobian of $\bby$, and
is known as the inverse magnification matrix. We choose coordinates such
that the origin of the source plane $(\bby = \bmo)$ is on the caustic.
In addition, we require that the origin of the lens plane $(\bx =
\bmo)$ maps to the origin of the source plane. We are interested in sources that lie near the caustic point ($\bby = \bmo$), which give rise to lensed images near the critical point ($\bx = \bmo$). In this case, we may expand the lens potential in a Taylor series about the point $\bx = \bmo$. We then find that the inverse magnification matrix at $\bx = \bmo$ is given by
\begin{equation}
  \left. \frac{\partial \bby}{\partial \bx} \right|_{\bmo} =
    \left[ \begin{array}{cc}
      1 - 2\ha & -\hb \\
      -\hb & 1 - 2\hc
    \end{array} \right],
\end{equation}
where
\begin{equation}
  \ha = \frac{1}{2} \psi_{11} (\bmo), \qquad
  \hb =  \psi_{12} (\bmo), \qquad
  \hc =  \frac{1}{2} \psi_{22} (\bmo).
\end{equation}
The subscripts indicate partial derivatives of $\psi$ with respect
to $\bx$.  Note that $\psi$ has no linear part (since $\bby=\bmo$
when $\bx=\bmo$). For $\bby = \bmo$ to be a caustic point, we must
have $(1-2\ha)(1-2\hc) - \hb^2 = 0$. In addition, at least one of
$(1-2\ha)$, $(1-2\hc)$, and $\hb^2$ must be non-zero
\citep[p.~349]{Petters_book}. Consequently, $(1-2\ha)$ and
$(1-2\hc)$ cannot both vanish. Without loss of generality, we assume
that $1-2\ha \neq 0$.

We now introduce the orthogonal matrix
\citep[see][p.~344]{Petters_book}
\begin{equation}
  {\bf M} = \frac{1}{\sqrt{(1 - 2\ha)^2 + \hb^2}}
    \left[ \begin{array}{cc}
      1 - 2\ha & -\hb \\
      \hb & 1 - 2\ha
    \end{array} \right] ,
\end{equation}
which diagonalizes $\left. \partial \bby / \partial \bx
\right|_{\bmo}$. We then define new orthogonal coordinates by
\begin{equation}
  \bt \equiv (\theta_1, \theta_2) \equiv  {\bf M}\, \bx ,
  \qquad
  \bu \equiv (u_1, u_2) \equiv {\bf M}\, \bby .
\end{equation}
Note that the coordinate changes are the {\em same} in the lens and
source planes. The advantage of using the same transformation in
both the lens and source planes is that the lens equation takes the
simple form
\begin{equation} \label{eq:lens-ortho}
  \bu = \bt - \grad\psi(\bt) ,
\end{equation}
and that the inverse magnification can be written as
\begin{equation} \label{eq:inv-mag-rot}
  \mu^{-1} = \det \left(\frac{\partial \bu}{\partial \bt}\right) .
\end{equation}
The old and new coordinate systems in the source and image planes
are shown in Figure \ref{fig:u-theta-def}. Since the caustic in the
source plane maps to the critical curve in the image plane, the origin
of the $(\theta_1, \theta_2)$ frame is determined from that of the
$(u_1, u_2)$ frame. The orientation of the $(\theta_1, \theta_2)$ axes
is determined by the matrix ${\bf M}$, and is not necessarily related
to the tangent to the critical curve.

Using the local orthogonal coordinates $\bu$ and $\bt$,
\citet[p.~346]{Petters_book} showed that $\bx = \bmo$ is a fold
critical point if and only if the following conditions hold
\begin{equation} \label{eq:eq-pot-fconstraints}
  (1-2\ha)(1-2\hc) = \hb^2 , \quad
  1-2\ha \neq 0 , \quad
  \psi_{222} (\bmo) \neq 0 .
\end{equation}
For a cusp, the third condition above is replaced by the
requirements that
\begin{equation} \label{eq:eq-pot-cconstraints}
  \psi_{222} (\bmo) = 0 , \quad
  \psi_{122} (\bmo) \neq 0 , \quad
  \psi_{2222} (\bmo) \neq 0 .
\end{equation}
Note in particular that $\psi_{222} (\bmo) = 0$ for a cusp while
$\psi_{222} (\bmo) \neq 0$ for a fold; this indicates that these two
cases must be treated separately.

We are interested in obtaining the positions,
magnifications and time delays of images near critical points. Since these
quantities depend only on the behavior of the lens potential near
a fold or cusp point, we can expand $\psi(\bt)$ in a Taylor series
about the point $\bt = \bmo$. To obtain all the quantities of
interest to leading order, we must expand the lens potential to
fourth order in $\bt$ \citep[pp.~346--347]{Petters_book}:
\begin{equation}
  \psi(\theta_1, \theta_2) =
      \frac{1}{2}(1-K)\,\theta_1^2
    + \frac{1}{2}\,\theta_2^2
    + e\,\theta_1^3
    + f\,\theta_1^2 \theta_2
    + g\,\theta_1   \theta_2^2
    + h\,           \theta_2^3
    + k\,\theta_1^4
    + m\,\theta_1^3 \theta_2
    + n\,\theta_1^2 \theta_2^2
    + p\,\theta_1   \theta_2^3
    + r\,           \theta_2^4 , \label{eq:psiexp}
\end{equation}
where the coefficients $\{K, e, f, g, \ldots, r\}$ are partial derivatives of the potential evaluated at the origin.
Lensing observables are independent of a constant term in the
potential, so we have not included one. Since $\bt=\bmo$ maps to
$\bu=\bmo$, any linear terms in the potential must vanish. In the
second order terms, the coefficients of the $\theta_1 \theta_2$ and
$\theta_2^2$ terms are set to $0$ and $1/2$, respectively, in order
to ensure that the point $\bt = 0$ is a critical point
\citep[see Appendix A1 of][]{Keeton-fold}.

\section{The Fold Case \label{sec:fold}}

In this section, we use perturbation theory
\citep[e.g.,][]{Bellman_pert} to derive an analytic relation
between the time delays in a fold pair. To derive this expression,
we must first obtain the image positions at which the time delay
is evaluated. These results were derived by \citet{Keeton-fold}.
We offer a summary of their analysis in Section \ref{sub:fold_image}
and present our new results for the time delay in Section
\ref{sub:fold_time}.

\subsection{Image Positions \label{sub:fold_image}}

Since we are considering a source near a fold point, we write its
position in terms of a scalar parameter $\epsilon$, which we take to
be small but finite. In particular, let $\bu \to \epsilon \bu$.
Combining equations~(\ref{eq:lens-ortho}) and (\ref{eq:psiexp}) we can
write the lens equation as
\begin{eqnarray}
   \epsilon u_1 &=& K\,\theta_1
    - \left( 3 e\,\theta_1^2 + 2 f\,\theta_1 \theta_2 + g\,\theta_2^2 \right)
    - \left( 4 k\,\theta_1^3 + 3 m\,\theta_1^2 \theta_2
        + 2 n\,\theta_1 \theta_2^2 + p\,\theta_2^3 \right) ,
    \label{eq:lens1-u1} \\
   \epsilon u_2 &=& - \left( f\,\theta_1^2 + 2 g\,\theta_1 \theta_2
    + 3 h\,\theta_2^2 \right) - \left( m\,\theta_1^3
    + 2 n\,\theta_1^2 \theta_2 + 3 p\,\theta_1 \theta_2^2
    + 4 r\,\theta_2^3 \right)
    \label{eq:lens1-u2}
\end{eqnarray}
\citep[see][Theorem 9.1]{Petters_book}. To find the image positions,
we expand $\theta_1$ and $\theta_2$ in a power series in $\epsilon$.
Since the left-hand sides of equations~(\ref{eq:lens1-u1}) and
(\ref{eq:lens1-u2}) are accurate to $\mathcal{O}(\epsilon)$, the
right-hand sides must be accurate to the same order.  Noting that
the lowest-order terms on the right-hand side are linear or quadratic
in $\bt$, we write
\begin{eqnarray}
  \theta_1 &=& \alpha_1\,\epsilon^{1/2} + \beta_1\,\epsilon + \order{\epsilon}{3/2} ,
    \label{eq:theta1-fold} \\
  \theta_2 &=& \alpha_2\,\epsilon^{1/2} + \beta_2\,\epsilon + \order{\epsilon}{3/2} .
    \label{eq:theta2-fold}
\end{eqnarray}
Substituting into the lens equation, we obtain
\begin{eqnarray}
  0 &=& (\alpha_1 K)\epsilon^{1/2} - (3 \alpha_1^2 e + 2 \alpha_1 \alpha_2 f
    + \alpha_2^2 g - \beta_1 K + u_1)\epsilon + \order{\epsilon}{3/2} ,
    \label{eq:lens2-u1} \\
  0 &=& -(\alpha_1^2 f + 2 \alpha_1 \alpha_2 g + 3 \alpha_2^2 h + u_2) \epsilon
    - \left[ 2 \alpha_1 \beta_1 f
    + 2 (\alpha_1 \beta_2 + \alpha_2 \beta_1) g
    + 6 \alpha_2 \beta_2
    h + \alpha_1^3 m + 2 \alpha_1^2 \alpha_2 n + 3 \alpha_1 \alpha_2^2 p
    + 4 \alpha_2^3 r \right] \epsilon^{3/2} \nonumber\\
  && + \order{\epsilon}{2} .
    \label{eq:lens2-u2}
\end{eqnarray}
Note that these equations are carried to different orders in
$\epsilon$, since the leading-order term in equation~(\ref{eq:lens1-u1})
is linear in $\bt$, while the leading-order term in equation~(\ref{eq:lens1-u2})
is quadratic in $\bt$.

Since $\epsilon$ is non-zero, equations~(\ref{eq:lens2-u1}) and
(\ref{eq:lens2-u2}) must be satisfied at each order in $\epsilon$.
We can work term by term to solve for the coefficients $\alpha_1$,
$\alpha_2$ and $\beta_2$, and then write the image positions as
\begin{eqnarray}
  \theta_1^{\pm} &=& \frac{3 h u_1 - g u_2}{3 h K}\ \epsilon + \order{\epsilon}{3/2} , \\
  \theta_2^{\pm} &=& \pm \sqrt{\frac{-u_2}{3h}}\ \epsilon^{1/2}
    - \frac{3 g h u_1 - g^2 u_2}{9 h^2 K}\ \epsilon + \order{\epsilon}{3/2} \,,
\end{eqnarray}
where the $\pm$ labels indicate the parities of the images. From these equations, we see that two images form near the point
$\bt=\bmo$ on the critical curve, provided that $(-u_2/3h) > 0$.
Since $h \le 0$ for standard lens potentials (e.g., an isothermal
ellipsoid or isothermal sphere with shear), we must have $u_2 > 0$.
In other words, the source must lie inside the caustic in order to
produce a pair of fold images. In practice, a more useful quantity
is the image separation, given by
\begin{equation} \label{eq:d1ser}
  d_1 = \sqrt{\frac{-4 u_2}{3h}}\ \epsilon^{1/2} + \order{\epsilon}{3/2} .
\end{equation}

\subsection{Time Delays \label{sub:fold_time}}

To find the time delay between the two fold images, we begin with
the general expression for the scaled time delay
\citep[e.g.,][]{Schneider_lensing}:
\begin{equation}
  \hat\tau (\bt) \equiv \frac{\tau (\bt)}{\tau_0}
    = \frac{1}{2} | \, \bt - \bu \, |^2 - \psi(\bt) .
\end{equation}
The scale factor is given by
\begin{equation}
  \tau_0 = \frac{1+z_L}{c} \frac{D_L D_S}{D_{LS}} ,
\end{equation}
where $D_L, D_S$ and $D_{LS}$ are the angular-diameter distances
from the observer to lens, observer to source, and lens to source,
respectively. The lens redshift is denoted by $z_L$. Making the
substitution $\bu \rightarrow (\epsilon u_1, \epsilon u_2)$, we have
for the two fold images
\begin{eqnarray}
  \hat\tau_- \ \equiv\ \hat\tau(\bt_-) &=& \ \, \sqrt{-\frac{4}{27 h}} \, (\epsilon u_2)^{3/2} + \order{\epsilon}{2} , \\
  \hat\tau_+ \ \equiv\ \hat\tau(\bt_+) &=& - \, \sqrt{-\frac{4}{27 h}} \, (\epsilon u_2)^{3/2} + \order{\epsilon}{2} .
\end{eqnarray}
The time delay between images is then (cf. \citealt{Schneider_lensing}, pp 190 - 191)
\begin{equation} \label{eqn:tfold-reln}
  \Delta \hat\tau_{\rm fold} \equiv \hat\tau_- - \hat\tau_+
  = \sqrt{-\frac{16}{27 h}} \, (\epsilon u_2)^{3/2} + \order{\epsilon}{2}
  = - \, \frac{h}{2} d_1^3 + \order{\epsilon}{2} ,
\end{equation}
which is positive, in agreement with the general result that images with negative parity trail those with positive parity. We find that the only coefficient from the lens potential that enters the expression for the differential time delay is the parameter $h
=\psi_{222}(\bmo)/6$. We also see that to leading order in
$\epsilon$, the image separation and the
differential time delay depend only on the $u_2$ component of the
source position. Unlike the image positions, our
expression for the time delay does not involve any of the
fourth-order terms in the potential. This is because
the time delay involves the potential directly, while the
image positions depend on
first derivatives of the potential. This means that all fourth-order
terms in the potential enter the time delay at
$\order{\epsilon}{2}$, while these same terms enter at
$\order{\epsilon}{3/2}$ in quantities involving derivatives.

To summarize, $\Delta \hat\tau_{\rm fold} \propto (\epsilon
u_2)^{3/2} \propto d_1^3$. For comparison, $R_{\rm fold} \propto
d_1$. Since $d_1$ is small, a violation of the ideal relation
$\Delta \hat\tau_{\rm fold} = 0$ is more likely to indicate the
presence of small-scale structure in the lens galaxy than would be
indicated by a non-zero value of $R_{\rm fold}$.

\begin{figure}
\begin{center}
\includegraphics[width=0.9\textwidth]{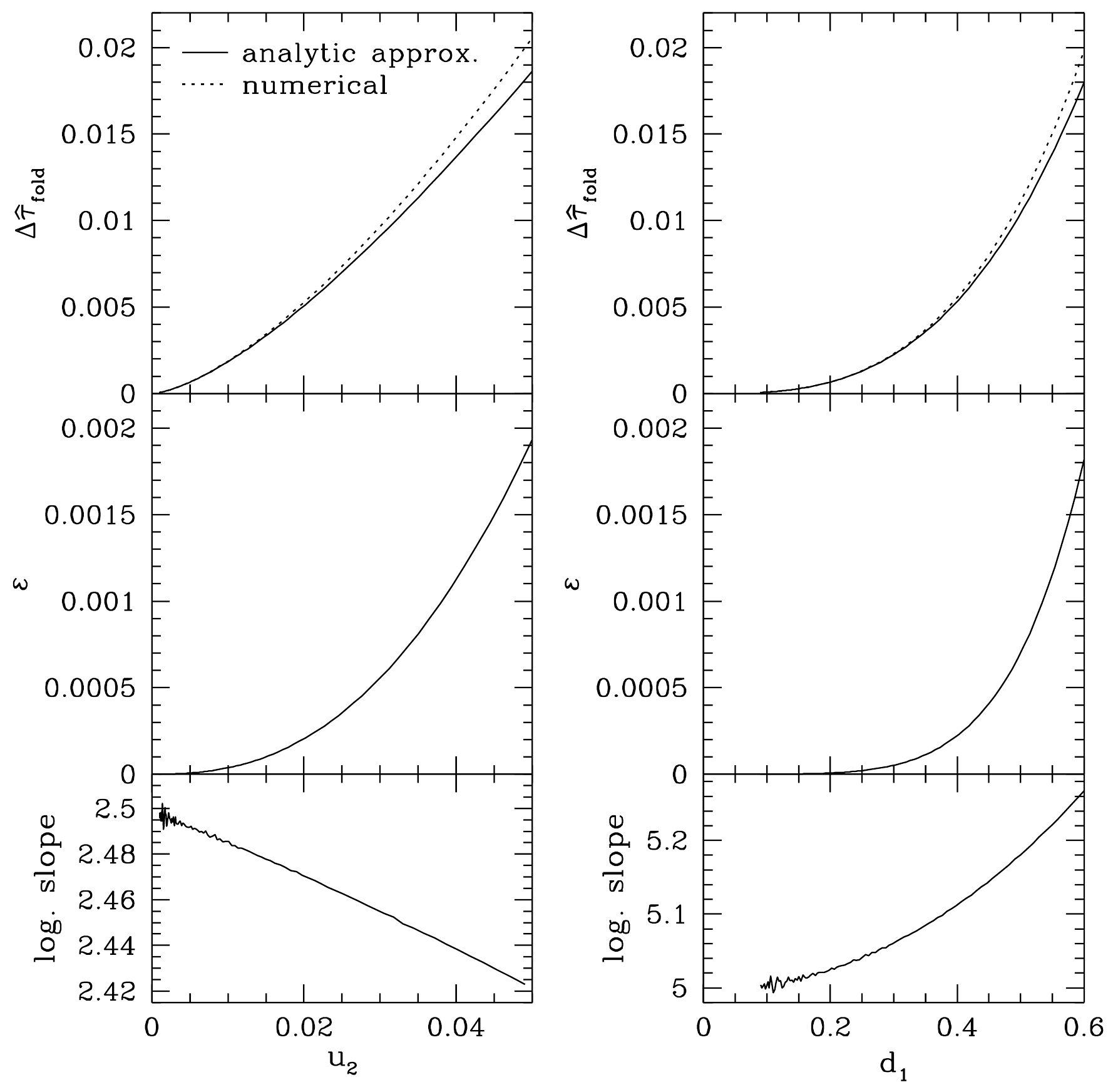}
\end{center}
\caption{
The top panels show the time delay for a fold pair as a function of
source position (left) and image separation (right), for an SIE lens
with axis ratio $q=0.5$. The solid line shows our analytic approximation
while the dotted line shows the exact result obtained by solving the
lens equation numerically. The quantities $u_2$ and $d_1$ are defined
in the text, and are shown here in units of the Einstein angle,
$\theta_E$.  The time delay is given in units of $\theta_E^2$.  The
middle panels show the error in the analytic approximation,
$\varepsilon = \Delta \hat\tau_{\rm fold}(\mbox{numerical}) -
\Delta \hat\tau_{\rm fold}(\mbox{analytic})$, due to our neglect of
higher-order terms.  The bottom panels show the logarithmic slope of
the error curve, $d(\ln \varepsilon)/d(\ln u_2)$ on the left and
$d(\ln \varepsilon)/d(\ln d_1)$ on the right.  There is some numerical
noise in the logarithmic slope due to numerical differentiation.  Together
the middle and bottom panels verify that our analytic expression is
accurate at order $\epsilon^{3/2}$ (cf.\ equation~\ref{eqn:tfold-reln}).
\label{fig:checkTfold}}
\end{figure}

Our analytic expression for $\Delta \hat\tau_{\rm fold}$ is only valid
for sources sufficiently close to the caustic that higher-order terms
are negligible. To quantify this statement, we compare our analytic
approximation with the differential time delay computed numerically
from the exact form of the lens equation. The numerical analysis requires a specific lens model, so we consider
an SIE with axis ratio $q=0.5$ as a representative example. We use the
software of \citet{Keeton-gravlens} to compute the astroid caustic,
and then choose a point on the caustic, far from a cusp, to serve as
the origin of the $(u_1,u_2)$ frame. For a given value of $u_2$, we
solve the exact lens equation to obtain the image positions and time
delay for the fold doublet. The top left panel of
Figure \ref{fig:checkTfold} shows $\Delta \hat\tau_{\rm fold}$ in
units of $\theta_E^2$ as a function of $u_2$ in units of $\theta_E$,
where $\theta_E$ is the Einstein angle of the lens.
The analytic and numerical results are in good agreement for sources
within $0.05 \theta_E$ of the caustic, although the curves do begin
to diverge as $u_2$ increases.  The difference between the numerical
and analytic curves (which represents the error in the analytic
approximation, denoted by $\varepsilon$) is shown in the middle left panel.
The bottom left panel shows the logarithmic slope of the error curve,
$d(\ln \varepsilon)/d(\ln u_2)$.  Together, the middle and bottom
left panels verify that our analytic approximation is accurate at
order $\epsilon^{3/2}$.  Furthermore, these panels suggest that
the next non-vanishing term is of order $\epsilon^{5/2}$ and has a
positive coefficient.  The interesting implication is that the
coefficient of the $\epsilon^2$ term seems to vanish.  At our current
order of approximation we are not able to determine whether this is
rigorously true, and if so, how general it is; we merely offer the
remark in the hope that it may be useful for future analytic studies.
For now, we focus on the verification that our analytic approximation
is accurate at the order to which we work.

Since the source position $u_2$ is not observable, we compare the
analytic and numerical time delays as a function of the image
separation $d_1$ in the right-hand panels of Figure \ref{fig:checkTfold}.
The range of $d_1$ corresponds to that used for $u_2$ in the
left-hand panels. For a canonical fold lens with
$d_1 = 0.46 \theta_E$ \citep{Keeton-fold}, our analytic expression
gives a very good approximation, indicating that our analysis can be
applied in astrophysically relevant situations---not just when the
source is asymptotically close to the caustic, but even when it lies
some small but finite distance away.

\section{The Cusp Case \label{sec:cusp}}
We now apply our perturbative method to the case of a source near a
cusp point. This approach has not been applied to cusp lenses
before. Appendix A of \citet{Keeton-cusp} derives the image positions
and magnifications for a cusp triplet assuming a simplified form of
the lens equation. As we noted in Section \ref{sec:intro}, this simplified
lens equation assumes that certain terms may be set to zero, using criteria
that are less than rigorous. We use the lens equation derived from
the fourth-order lens potential, and use perturbation theory to
verify the results of \citet{Keeton-cusp} and extend the analysis to a higher order of approximation. We also study time delays for a cusp lens for the first time. Our analysis does not involve simplifying assumptions, and indicates that perturbation theory is a powerful method in the study of lensing.

\subsection{Image Positions \label{sub:cusp_image}}

We again expand the image positions, magnifications and time delays
in the parameter $\epsilon$, but with one notable difference. For a
cusp oriented in the $u_1$ direction, a small ``horizontal''
displacement of $\epsilon u_1$ from the cusp point permits a
``vertical'' displacement of only $\epsilon^{3/2} u_2$ (see Fig.
\ref{fig:u-theta-def}), since larger vertical displacements would
imply a source position outside the caustic
\citep{Blandford_Narayan}. The lens equation is then
\begin{eqnarray}
   \epsilon u_1 &=& K\,\theta_1
    - \left( 3 e\,\theta_1^2 + 2 f\,\theta_1 \theta_2 + g\,\theta_2^2 \right)
    - \left( 4 k\,\theta_1^3 + 3 m\,\theta_1^2 \theta_2
           + 2 n\,\theta_1 \theta_2^2 + p\,\theta_2^3 \right) ,
    \label{eq:lens1-u1-cusp} \\
   \epsilon^{3/2} u_2 &=& - \left( f\,\theta_1^2 + 2 g\,\theta_1 \theta_2
    \right) - \left( m\,\theta_1^3
    + 2 n\,\theta_1^2 \theta_2 + 3 p\,\theta_1 \theta_2^2
    + 4 r\,\theta_2^3 \right)
    \label{eq:lens1-u2-cusp}
\end{eqnarray}
\citep[see][Theorem 9.1]{Petters_book}, where the $\theta_2^2$  term
of equation~(\ref{eq:lens1-u2}) does not appear, since $\psi_{222}(\bmo) = 0$
for a cusp, corresponding to $h = 0$ in equation~(\ref{eq:psiexp}).

As before, we write the image positions as a series expansion in
$\epsilon$, but now keeping an additional term (i.e., $\gamma_i
\epsilon^{3/2}$). This is necessary since the vertical component of
the source position enters the lens equation as $\epsilon^{3/2}
u_2$, rather than $\epsilon u_2$ as in the fold case. We have
\begin{eqnarray}
  \theta_1 &=& \alpha_1\,\epsilon^{1/2} + \beta_1\,\epsilon + \gamma_1\,\epsilon^{3/2} + \order{\epsilon}{2},
    \label{eq:theta1-cusp} \\
  \theta_2 &=& \alpha_2\,\epsilon^{1/2} + \beta_2\,\epsilon + \gamma_2\,\epsilon^{3/2} + \order{\epsilon}{2}.
    \label{eq:theta2-cusp}
\end{eqnarray}
The lens equation then becomes
\begin{eqnarray}
  0 &=& \alpha_1 K \epsilon^{1/2} - \left(3 e \alpha_1^2 + 2 f \alpha_1 \alpha_2
    + g \alpha_2^2  - \beta_1 K + u_1 \right)\epsilon \nonumber\\
  && - \left[4k\alpha_1^3 + 3m\alpha_1^2 \alpha_2 + p\alpha_2^3 +
      2\alpha_1(n\alpha_2^2 + 3e\beta_1 + f\beta_2)
    + 2\alpha_2(f\beta_1 + g\beta_2)-K\gamma_1\right]\epsilon^{3/2}
    + \order{\epsilon}{2} ,
    \label{eq:lens2-u1-cusp} \\
  0 &=& -(f \alpha_1^2 + 2g \alpha_1 \alpha_2)\epsilon - \left(u_2
    + m \alpha_1^3 + 2n\alpha_1^2 \alpha_2 + 3p\alpha_1 \alpha_2^2 +4 r
    \alpha_2^3 + 2f \alpha_1 \beta_1 +2 g \alpha_2 \beta_1 + 2g
    \alpha_1 \beta_2 \right)\epsilon^{3/2} \nonumber \\
  && - \left\{\beta_1 (f\beta_1 + 2g \beta_2)
    + \alpha_1^2(3m\beta_1 +2 n \beta_2) + 3
    \alpha_2^2(p\beta_1 + 4 r \beta_2) + 2g \alpha_2 \gamma_1
    + 2 \alpha_1 \left[\alpha_2\left(2n\beta_1 + 3p\beta_2\right) + f\gamma_1
    + g\gamma_2\right]\right\} \epsilon^2 \nonumber\\
  && + \order{\epsilon}{5/2} .
    \label{eq:lens2-u2-cusp}
\end{eqnarray}
As in the fold case (see equation~\ref{eq:lens2-u1}), we find
$\alpha_1 = 0$. Note that $\gamma_2$ appears only in the
$\epsilon^2$ coefficient of equation~(\ref{eq:lens2-u2-cusp}), but
in a term multiplied by $\alpha_1$. Hence, it will not be possible
to solve for $\gamma_2$. Fortunately, it turns out that the
expressions we would like to derive do not involve this parameter.

We can now write the lens equation as
\begin{eqnarray}
  0 &=& - \left(g \alpha_2^2 - \beta_1 K + u_1 \right)\epsilon - \left[
    p\alpha_2^3 +2\alpha_2(f\beta_1 + g\beta_2)-K\gamma_1\right]\epsilon^{3/2}
    + \order{\epsilon}{2} ,
    \label{eq:lens3-u1-cusp} \\
  0 &=& - \left(u_2 +4 r  \alpha_2^3 +2 g \alpha_2 \beta_1 \right)\epsilon^{3/2}
    - \left[\beta_1 (f\beta_1 + 2g \beta_2) +  3\alpha_2^2(p\beta_1
    + 4 r \beta_2) + 2g \alpha_2 \gamma_1  \right] \epsilon^2
    + \order{\epsilon}{5/2} .
    \label{eq:lens3-u2-cusp}
\end{eqnarray}
Recalling that these equations must be satisfied at each order in
$\epsilon$, we then solve for the unknown coefficients:
\begin{eqnarray}
  \beta_1  &=&  \frac{g \alpha_2^2 + u_1}{K} , \\
  \beta_2  &=& -\frac{(5 f g^2 + 5 g K p) \alpha_2^4
   + (6 f g u_1 + 3 K p u_1) \alpha_2^2 + f u_1^2}
   {2K \left(3 g^2 \alpha_2^2 + 6 K r \alpha_2^2 + g u_1 \right)} , \\
  \gamma_1 &=& \frac{(f g^3 - 2 g^2 K p + 12 f g K r + 6 K^2 p r) \alpha_2^5
   + (2 f g^2 u_1 - 2 g K p u_1 + 12 f K r u_1) \alpha_2^3 + f g u_1^2 \alpha_2}
   {K^2 \left(3 g^2 \alpha_2^2 + 6 K r \alpha_2^2 + g u_1 \right)} ,
\end{eqnarray}
where $\alpha_2$ satisfies the cubic equation
\begin{eqnarray}
  \alpha_2^3 + \frac{g u_1}{2 K r + g^2} \alpha_2 + \frac{K u_2}{4 K r + 2 g^2} = 0 .
\end{eqnarray}
This is equivalent to equation (A8) from \citet{Keeton-cusp}, after
making the replacements $\alpha_2 \rightarrow z, \, K \rightarrow c,
\, g \rightarrow -b/2, \, r \rightarrow -a/4$. To leading order, the
image positions can be written as
\begin{eqnarray}
  \theta_1 &=& \beta_1 \epsilon = \frac{g \alpha_2^2 + u_1}{K} \epsilon , \\
  \theta_2 &=& \alpha_2 \epsilon^{1/2} ,
\end{eqnarray}
which are equivalent to equation (A7) from \citet{Keeton-cusp}. The
distance between any two images $i$ and $j$ is then
\begin{equation}
  d_{ij} = [\alpha_2^{(i)} - \alpha_2^{(j)}] \epsilon^{1/2} + \order{\epsilon} \,.
\end{equation}

\subsection{Magnifications \label{sub:cusp_mag}}

\begin{figure}
\begin{center}
\includegraphics[width=0.9\textwidth]{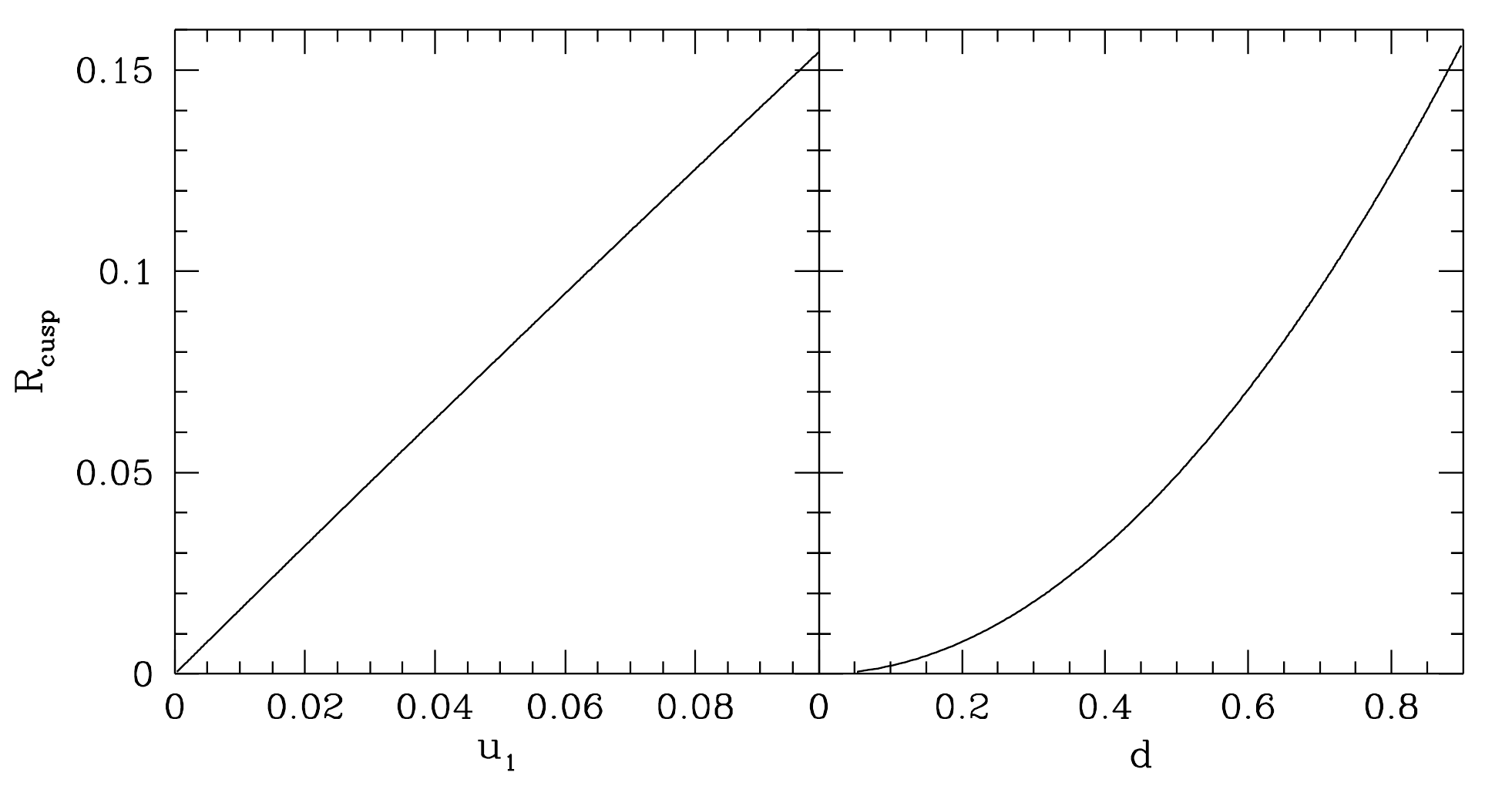}
\end{center}
\caption{
$\Rcusp$ as a function of $u_1$ (left) and $d$ (right) for an SIE with
$q=0.5$, obtained by solving the lens equation numerically.
\label{fig:checkRcusp}}
\end{figure}

Substituting our perturbative expressions for the image positions into
equation (\ref{eq:inv-mag-rot}), we find that the inverse magnification of a cusp image is given by:
\begin{equation}
  \mu^{-1} = -2[g u_1 + 3(g^2 + 2 K r) \alpha_2^2] \, \epsilon
    + [(24 f r -12 g p) \alpha_2^3 + (6 K p - 4 f g) \alpha_2 \beta_1
    - (8 g^2 + 24 K r) \alpha_2 \beta_2 - g K \gamma_1] \, \epsilon^{3/2}
    + \order{\epsilon}{2} .
\end{equation}
We then find from equation~(\ref{eq:fluxratdef}) that
\begin{equation}
  R_{\rm cusp} = 0 + \orderof{\epsilon} \approx \Acusp d^2,
\end{equation}
where $d = \max_{ij} d_{ij}$ is the largest separation between any two of the three cusp images. This expression shows that correction terms to the ideal cusp relation enter at second order in the image separation, which agrees with the conjecture of \citet{Keeton-cusp}. To see this, we define $m_i \equiv |\mu_i^{-1}|$, which allows us to write
\begin{equation}
  \Rcusp = \frac{m_B m_C - m_A m_C + m_A m_B}{m_B m_C + m_A m_C + m_A m_B} \,.
\end{equation}
If the leading-order term in the numerator vanishes, so does the leading-order term in $\Rcusp$. The zeroth-order term in $\Rcusp$ corresponds to a term of $\order{\epsilon}{2}$ in the numerator, since the leading-order term in the denominator is $\order{\epsilon}{2}$. By substituting the solutions for $\alpha_2$ into the numerator, we find that $\Rcusp = 0$ to lowest order, in agreement with \citet{Keeton-cusp}. We repeat this procedure for the next-leading term of $\order{\epsilon}{5/2}$ in the numerator, and find that $\Rcusp = 0$ at linear order in $d$ [i.e., $\order{\epsilon}{1/2}$] as well; this result was unattainable using the formalism of \citet{Keeton-cusp}.
To proceed to higher order, we must extend our perturbative analysis by including terms of the form $\delta_i \epsilon^2$ in equations (\ref{eq:theta1-cusp}) and (\ref{eq:theta2-cusp}) for the image positions. We denote the coefficient of $d^2$ by $\Acusp$, which we do not here write down, since that would require several pages. Given the complexity of this term, it is not practical to evaluate $\Acusp$ analytically. However, we have numerically computed $\Rcusp$ for the case of an SIE model with $q=0.5$ (see Fig. \ref{fig:checkRcusp}) and find that $\Rcusp \propto u_1 \propto d^2$. While it is conceivable that $\Acusp$ might be zero for some specific lens model, clearly this is not the case in general. We have thus demonstrated analytically that $\Rcusp$ vanishes through linear order in $d$, placing the numerical result of \citet{Keeton-cusp} on solid mathematical ground.

\subsection{Time Delays \label{sub:cusp_time}}

For the cusp case, the scaled time delay takes the form
\begin{equation}
  \hat\tau = \frac{1}{2} \left[ (\theta_1 - \epsilon u_1)^2 + (\theta_2 - \epsilon^{3/2} u_2)^2 \right]
    - \psi(\theta_1, \theta_2) .
\end{equation}
We find
\begin{equation}
  \hat\tau = \frac{1}{2 K} \left[(3 g^2 + 6 K r) \alpha_2^4 + 2 g u_1 \alpha_2^2
    + (K - 1) u_1^2 \right] \epsilon^2 + \order{\epsilon}{5/2} ,
\end{equation}
corresponding to a differential time delay of
\begin{equation}
  \Delta \hat\tau_{\rm cusp}^{(ij)} = \frac{1}{4 K} \left[ 2g \left(\alpha_2^{(i)} + \alpha_2^{(j)}\right) u_1 + 3 K u_2\right] \left(\alpha_2^{(j)} - \alpha_2^{(i)}\right) \epsilon^2 .
\end{equation}
Unlike the fold case, the time delay for a pair of cusp images
depends on both source coordinates $(u_1 , u_2)$. This means that it
is not possible to write our current expression strictly in terms of
observables, such as the image separation. Instead, all we can say
is that the time delay scales quadratically with $\epsilon$, or
alternatively, with the fourth power of the image separation.

In the fold case, we found that the time delay scales as
$\epsilon^{3/2}$ and only depends on the lens potential through the
parameter $h$. For a cusp, however, $h = 0$, so it is not surprising
that the lowest-order term in the time delay is of
$\order{\epsilon}{2}$. Furthermore, if we had not included the
$\gamma_i \, \epsilon^{3/2}$ terms in our expansions of the image
positions for a cusp [equations~(\ref{eq:theta1-cusp}] and
[\ref{eq:theta2-cusp})], it would not have been possible to obtain a
perturbative expression for the time delay; instead, we would simply
have found $\hat\tau = 0 + \order{\epsilon}{2}$.

\section{Summary}

We have developed a unified, rigorous framework for studying lensing near fold and cusp critical points, which can (in principle) be extended to arbitrary order. We have found that the differential time delay of a fold pair assumes a particularly simple form, depending only on the image separation and the Taylor coefficient $h = \psi_{222} (\bmo)/6$. This result is astrophysically relevant, since it is quite accurate even for sources that are not asymptotically close to the caustic. We have also obtained perturbative expressions for the image positions, magnifications and time delays of a cusp triplet. These results rest on the key insight that a source at a given distance $\epsilon$ from a cusp along the relevant symmetry axis of the caustic can only move a perpendicular distance of $\epsilon^{3/2}$ in order to remain inside the caustic \citep{Blandford_Narayan}. We have shown rigorously that the distance dependence of the magnification ratio $\Rcusp$ conjectured by \citet{Keeton-cusp} is correct. We have also demonstrated that the leading-order expression for the image positions is given by the relations presented by \citet{Keeton-cusp}, and have provided the necessary framework for deriving the image positions corresponding to a Taylor expansion of the lens potential at arbitrary truncation order. Finally, we have derived cusp time delays analytically for the first time.
Our results provide a rigorous foundation for identifying anomalous flux ratios and time delays in gravitational
lens systems, and for using them to study small-scale structure in the
mass distributions of distant galaxies.

\section*{Acknowledgments}

We thank A. O. Petters and Peter Schneider for helpful discussions
and suggestions, and for their careful reading of the manuscript.
We also thank the anonymous referee for useful comments.


\end{document}